\documentclass[journal]{IEEEtran}

\usepackage{graphicx}
\usepackage{stfloats}
\usepackage[flushleft,footnotesize]{caption2}
\usepackage{amsmath}
\usepackage[numbers,sort&compress]{natbib}

\begin{document}
\bibliographystyle{iopart-num}
\title{A Low-Fluorine Solution with the F/Ba Mole Ratio of 2 for the Fabrication of YBCO Films}

\author{\IEEEauthorblockN{Wei Wu\IEEEauthorrefmark{1}\IEEEauthorrefmark{2}, Feng Feng\IEEEauthorrefmark{3}\IEEEauthorrefmark{1}, Yue Zhao\IEEEauthorrefmark{2}, Xiao Tang\IEEEauthorrefmark{2}, Yunran Xue\IEEEauthorrefmark{4}\IEEEauthorrefmark{1}, Kai Shi\IEEEauthorrefmark{1}, Rongxia Huang\IEEEauthorrefmark{1}, Timing Qu\IEEEauthorrefmark{4}\IEEEauthorrefmark{3}, Xiaohao Wang\IEEEauthorrefmark{3}, Zhenghe Han\IEEEauthorrefmark{1} and Jean-Claude Grivel\IEEEauthorrefmark{2}}

\IEEEauthorblockA{\IEEEauthorrefmark{1}Applied Superconductivity Research Center, Department of Physics, Tsinghua University, Beijing 100084, China}

\IEEEauthorblockA{\IEEEauthorrefmark{2}Department of Energy Conversion and Storage, Technical University of Denmark, Frederiksborgvej 399, DK-4000 Roskilde, Denmark}

\IEEEauthorblockA{\IEEEauthorrefmark{3}Division of Advanced Manufacturing, Graduate School at Shenzhen, Tsinghua University, Shenzhen 518055, China}

\IEEEauthorblockA{\IEEEauthorrefmark{4}Department of Mechanical Engineering, Tsinghua University, Key Laboratory for Advanced Materials Processing Technology, Ministry of Education, Beijing 100084, China}

\thanks{Corresponding author: Feng Feng (Email:feng.feng@sz.tsinghua.edu.cn)}}

\maketitle

\begin{abstract}
In the reported low-fluorine MOD-YBCO studies, the lowest F/Ba mole ratio of the precursor solution was 4.5. However, further lowering the F/Ba ratio is important according to the researches of YBCO thick film. On the other hand, the F/Ba ratio is necessary to be at least 2 for the full conversion of the Ba precursor to BaF$_2$ to avoid the formation of BaCO$_3$, which is detrimental to the superconducting performance. In this study, a novel solution with the F/Ba mole ratio of 2 was developed, in which the fluorine content was only about 10.3\% of that used in the conventional TFA-MOD method. Attenuated total reflectance-Fourier transformed-infrared spectra(ATR-FT-IR) revealed that BaCO$_3$ was remarkably suppressed in the as-pyrolyzed film and eliminated at 700 $^\circ$C. Thus YBCO films with a critical current density ($J_\mathrm{c}$) over 5 MA cm$^{-2}$ (77 K, 0 T, 200 nm thickness) could be obtained on LAO single crystal substrates. In-situ FT-IR spectra showed that no obvious fluorinated gaseous by-products were detected in the pyrolysis step, which indicated that all of the F atoms might remain in the film as fluorides. X-ray diffraction (XRD) $\theta$/2$\theta$ scan showed that BaF$_2$, but neither YF$_3$ nor CuF$_2$, was detected in the films quenched at 400 - 800 $^\circ$C. The formation priority of BaF$_2$ over YF$_3$ and CuF$_2$ was interpreted by the chemical equilibrium of the potential reactions. Our study could enlarge the synthesis window of the precursor solution for MOD-YBCO fabrication and open a gate to study the fluorine content in the precursor solution continuously and systematically.
\end{abstract}


\section{Introduction}

Trifluoroacetate metal organic deposition (TFA-MOD), which was first reported by Gupta \itshape et al \upshape \cite{Gupta1988}, is one of the most popular methods to fabricate high-performance YBa$_2$Cu$_3$O$_{7-\delta}$ (YBCO) superconducting films \cite{MCINTYRE1990JAP,Smith1999IEEE,Obradors2006SuST,Araki2003SuST}. The fluorine content in the precursor solution of the TFA-MOD method is of great significance, because the formation of BaCO$_3$, which is very stable during the heat treatment and will deteriorate the critical current density ($J_\mathrm{c}$) of the fully processed YBCO films, can be avoided by the formation of BaF$_2$ \cite{Gupta1988, MCINTYRE1990JAP}; then the fluorine content of BaF$_2$ can be easily removed by the generation and release of HF gas \cite{Araki2001}.

However, the conventional TFA-MOD method requires a pyrolysis step that usually takes more than 10 h to complete \cite{Araki2003SuST}. Such a slow process constitutes a serious barrier for industrial production. Lowering the fluorine content in the precursor solution is considered to be effective in shortening the time requirement during the pyrolysis step \cite{Tokunaga2004PC}. This idea has been realized in many groups by substituting fluorine-free salt(s) for one or two specific TFA salt(s) in the precursor solution. To the best of our knowledge, there were mainly three routes of metal salt substitution(s): (1) Cu salt \cite{Tokunaga2004PC}; (2) Cu and Y salts \cite{Xu2005,Nakaoka2007}; (3) Cu and Ba salts \cite{Chen2012SuST}. Relative to the conventional TFA-MOD solution (100\% fluorine content), the fluorine contents in these precursor solutions could be estimated to be about 53.8\%, 30.8\% (or 23.1\% if the poor-Ba stoichiometry is used \cite{Ichikawa2009PC}) and 23.1\%, respectively.


Using a solution with lower fluorine content to prepare YBCO films would be beneficial. First, such a study could help to reduce the environmentally harmful fluorine by-products. Second, lowering the fluorine content could lead to a lower F/Ba ratio in the as-pyrolyzed films, which might be profound in the YBCO thick film fabrication. In the multi-coated MOD-YBCO thick film research by Li \itshape et al \upshape \cite{Li2007IEEE} and Feenstra \itshape et al \upshape \cite{Feenstra2009IEEE}, a modification for lowering the F/Ba ratio prior to the crystallization step was proposed, based on the hypothesis of transient liquid phase during the crystallization \cite{Yoshizumi2004PC,Wesolowski2006PC}. In their studies, lowering the F/Ba ratio was achieved by altering the water partial pressure and heating rate \cite{Li2007IEEE}, or by inserting a medium-temperature anneal ($<$650 $^\circ$C, 60 min) named as "F-Module" between pyrolysis and crystallization \cite{Feenstra2009IEEE}.


In order to develop a new solution with lower fluorine content, two facts should be noticed in the conventional TFA-MOD method \cite{Gazquez2006,Zalamova2010SuST} and the low-fluorine MOD studies \cite{Chen2012SuST,Matsuda2008JMR,Armenio2011}: (1) The phases involved in the formation of YBCO are CuO, BaF$_2$ (or its partially oxidized phase) and Y$_2$Cu$_2$O$_5$ (or Y$_2$O$_3$). This fact is also found in the ex-situ "BaF$_2$ process" \cite{Wong-Ng2004SuST}, which was considered to be the origin of the TFA-MOD method \cite{Araki2003SuST}. It is clear that the Y and Cu are not required to be fluorinated during the formation of YBCO. (2) In the as-pyrolyzed films, the fluorine containing phase is the solid solution of BaF$_2$ and YF$_3$. Then during the intermediate phase evolution before YBCO formation, YF$_3$ will convert to oxide, indicating its stability is lower than BaF$_2$. Thus BaF$_2$ may form prior to YF$_3$ and a starting fluorine content in the precursor solution with the F/Ba mole ratio of 2 might be adequate for the full conversion of the Ba precursor to BaF$_2$.



In the present work, a precursor solution with the F/Ba mole ratio of 2 was developed and thus named after "F/Ba-2". The fluorine content in this solution is only about 10.3\% of that used in the conventional TFA-MOD solution, which is much lower than that of any reported low-fluorine studies. YBCO thin films with favorable superconducting performance could be obtained repeatedly using this solution. The behaviour of the precursor during the heat treatment was also studied by multiple characterization methods and the analyses of chemical equilibrium.


\section{Experimental details}
\subsection{Sample preparation}

To synthesize the solution F/Ba-2, Y, Ba and Cu acetates with a stoichiometric ratio of 1:2:3, trifluoroacetic acid (TFA), deionized water, and propionic acid were mixed directly. The quantity of TFA was 10.3 mol.\% of the total CH$_3$COO$^{-}$ anion with a uncertainty estimated to be 0.5\%. After stirring for 1 h, the obtained solution was refined in a BUCHI Rotavapor R-210 rotary evaporator under decompression for 2 h. Methanol and ammonia water were then added to the gel-like production to obtain the precursor solution. Details of the reagents used to synthesize the solution F/Ba-2, which had a molar concentration of total metal ions of 1.2 mol L$^{-1}$, are summarized in Table 1. Polyethylene glycol (PEG) 1500, which can help to alleviate stress generation and prevent buckling during a rapid pyrolysis step \cite{Wu2013}, was used as an additive in the precursor solution. To obtain thicker films, a solution F/Ba-2 with a concentration of 1.6 mol L$^{-1}$ was also prepared. For convenience, these two solutions were named after "F/Ba-2-1.2M" and "F/Ba-2-1.6M", respectively. Unless expressly noted, the results in this paper is based on the solution F/Ba-2-1.2M.

\begin{table}[htbp]\label{table1}
\tabcolsep 0pt
\centering
\caption{The detailed reagent quantities utilized to synthesize the solution F/Ba-2 (20 mL, 1.2 mol L$^{-1}$).} \vspace*{-12pt}
\begin{center}
\def\temptablewidth{0.45\textwidth}
{\rule{\temptablewidth}{1pt}}
\begin{tabular*}{\temptablewidth}{@{\extracolsep{\fill}}ccccccc}
Reagent & Quantity \\
\hline
Y(CH$_3$COO)$_3$$\cdot$4H$_2$O (Sigma-Aldrich) & 0.004 mol \\
Ba(CH$_3$COO)$_2$ (Alfa Aesar) & 0.008 mol \\
Cu(CH$_3$COO)$_2$$\cdot$H$_2$O (Alfa Aesar) & 0.012 mol \\
CF$_3$COOH (Alfa Aesar) & 0.0054 $\pm$ 0.0003 mol \\
CH$_3$CH$_2$COOH (Alfa Aesar) & excess \\
CH$_3$OH (Sigma-Aldrich) & 10 mL \\
NH$_3$$\cdot$H$_2$O (Alfa Aesar) & 3.3 mL \\
       \end{tabular*}
       {\rule{\temptablewidth}{1pt}}
       \end{center}
       \end{table}

\begin{figure}[h]
\flushleft
\includegraphics[width=0.48\textwidth]{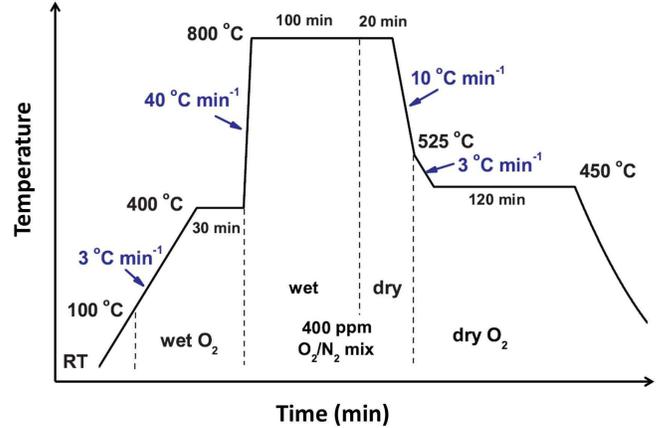}
\caption{The entire heat treatment process utilized to fabricate YBCO films in this study.} \label{HT}
\end{figure}

The precursor solution was coated on lanthanum aluminate (LAO) single crystal substrates (5$\times$5 mm$^2$) by a spin coater. The rotation speed and acceleration time were 8000 rpm and 1 s, respectively. Heat treatment was conducted in a MTI GSL-1500X tube furnace. First, the samples were pyrolyzed starting at room temperature (RT) to 400 $^{\circ}$C at a heating rate of 3 $^{\circ}$C min$^{-1}$ in a humid O$_2$ gas atmosphere. Then the pyrolyzed samples were crystallized at 800 $^{\circ}$C for 2 h in a 400 ppm O$_2$/N$_2$ atmospheric mixture. The YBCO superconducting thin films could then be obtained after 2 h of annealing in a dry O$_2$ gas atmosphere. The entire heat treatment process is illustrated in Figure \ref{HT}. To trace the chemical reaction path between the pyrolysis and the crystallization step, a series of samples were quenched at 400-800 $^{\circ}$C in this profile.

\subsection{Characterization and measurements}

Optical microscopy (OM) observation was conducted on the pyrolyzed films with a Nikon ME600 Polarized Optical Microscopy under the condition of bright field. The quality of the pyrolyzed films were evaluated by the OM observation. Scanning electron microscopy (SEM) characterization was performed in LEO 1530 with an in-lens detector (EHT = 10 keV). Surface morphologies of the YBCO films could be observed, and the film thickness was estimated according to the SEM cross sectional images.

Attenuated total reflectance - Fourier transformed-infrared spectra (ATR-FT-IR) were taken with a Perkin¨CElmer Spectrum GX FTIR system to detect the presence of BaCO$_3$. This sensitive technique allows the detection of BaCO$_3$ at a level as low as 2 wt\% \cite{Vermeir2009SuST}.

Differential scanning calorimetry (DSC) and thermogravimetric analysis (TGA) were carried out using a Netzsch STA 449C thermal analyzer in a humid oxygen atmosphere at temperatures ranging from RT to 400 $^{\circ}$C and a heating rate of 3 $^{\circ}$C min$^{-1}$. Samples for the DSC-TGA measurement were in the form of bulk gels, which were prepared by slowly evaporating the free solvent in the precursor solution at 70 $^{\circ}$C for 12 h. FT-IR spectra of the gaseous by-products generated in the decomposition reactions were recorded simultaneously by a BRUKER Tensor 27 FT-IR spectrometer connected to the thermal analyzer. The background carbon dioxide signal was subtracted. Considering the presence of water in the atmosphere, water signals not only from the background but also from the product were always subtracted to reveal signals from other gaseous phases.


The identification of the phases present in the films quenched at 400-800 $^\circ$C and in the fully processed YBCO films were performed with X-ray diffraction (XRD) $\theta/$2$\theta$ scans in a BRUKER D8 diffractometer, which used a Cu $K_\alpha$ source and a four-circle sample holder. The scan speed was 3 s/step with an increment of 0.02 $^{\circ}$/step. In our study, it should be noted that when the alignment of sample was perfect, many undesirable peaks of the LAO single crystal substrate due to the X-ray source impurity would appear. To avoid them and therefore uncover the relatively weak signals from some intermediate phases, the XRD $\theta/$2$\theta$ scans were always conducted while the sample was slightly deviated from its perfect alignment situation. XRD $\phi$ scan was utilized to study the in-plane texture of the highly aligned phases in the films. The scan speed was 0.6 s/step with an increment of 0.02 $^{\circ}$/step.



The superconducting properties of YBCO films were measured in a CRYOGENIC cryogen-free measuring system. Critical temperature ($T_\mathrm{c}$) was determined by the AC susceptibility measurement with the magenetic field amplitude of 0.1 mT and frequency of 21 Hz. Magnetization hysteresis loops of YBCO films were recorded by a vibrating sample magnetometer (VSM) at 77 K, under a magnetic field perpendicular to the surface of the LAO substrate. The critical current density ($J_\mathrm{c}$) was calculated using equation (\ref{Jc-equation}), which is based on the extended Bean critical state model \cite{GYORGY1989APL}:

\begin{equation}\label{Jc-equation}
  J_\mathrm{c} = \frac{2\Delta m}{Va(1-a/3b)}
\end{equation}

where a, b, V and $\Delta m$ are the width, length, volume of the film (a $\leq$ b) and the opening of the magnetization hysteresis loop at a certain magnetic field, respectively.

\section{Results}
\subsection{Feasibility study of YBCO film fabrication}

\begin{figure}[h]
\centering
\includegraphics[width=0.35\textwidth]{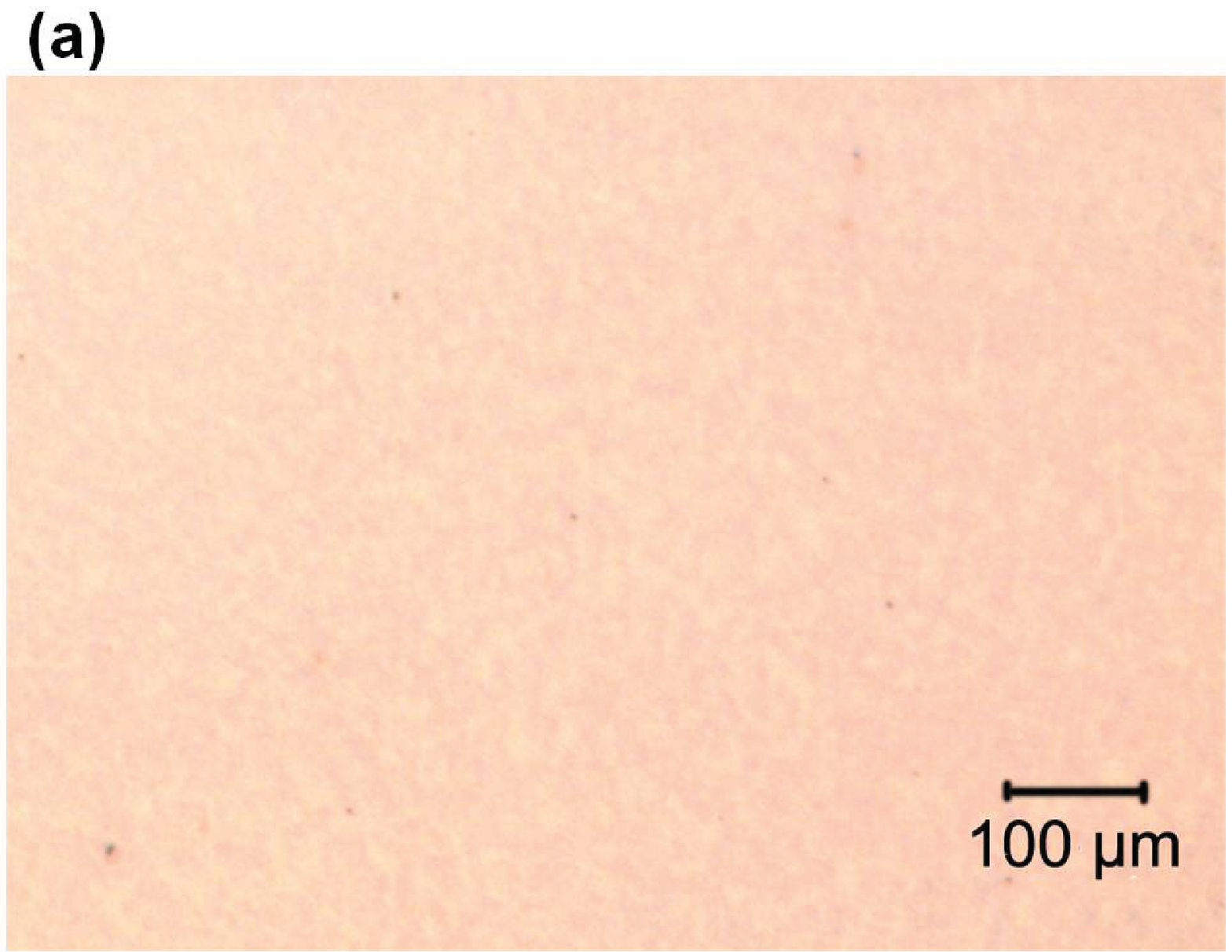}
\flushleft
\includegraphics[width=0.48\textwidth]{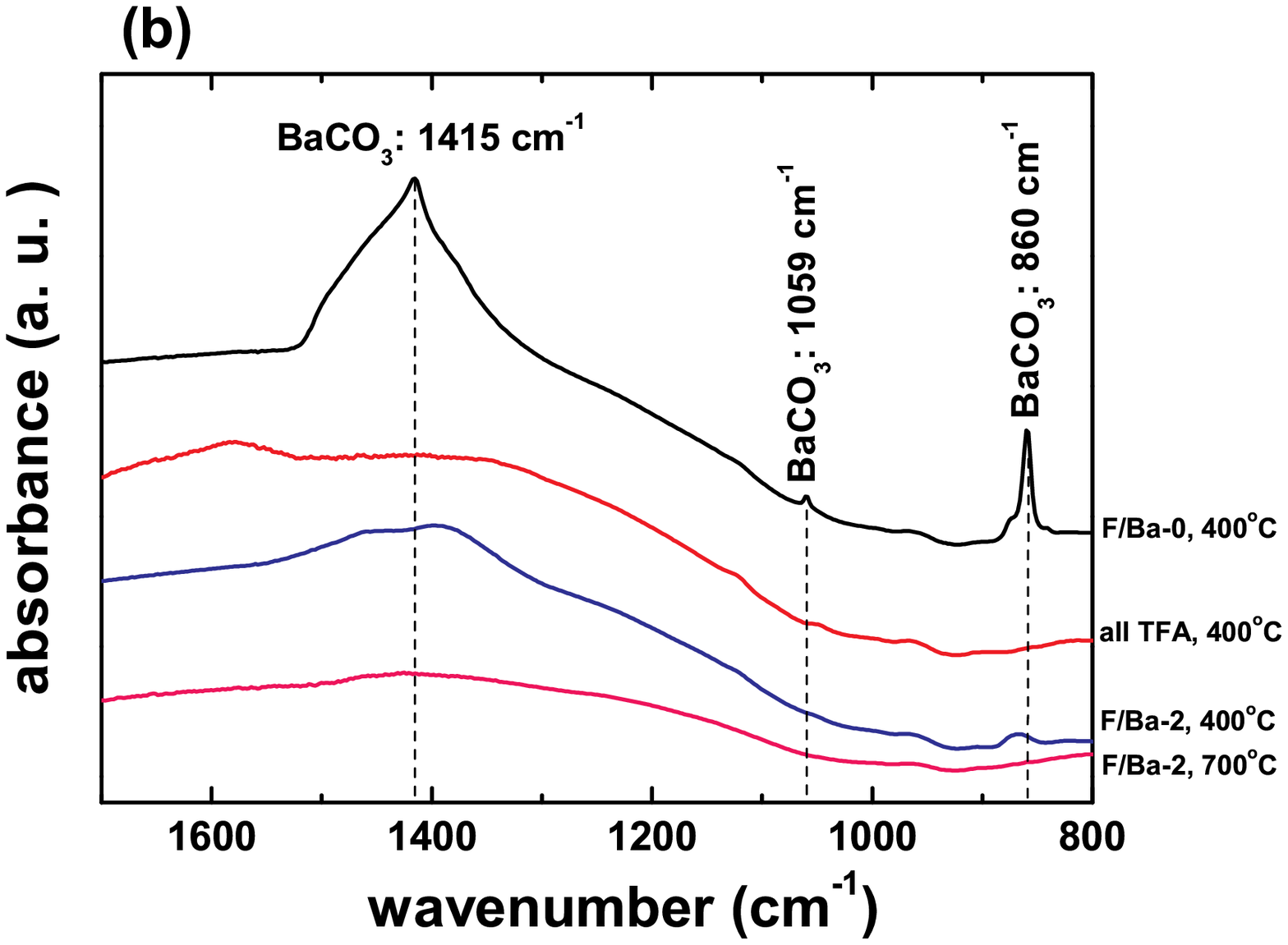}
\caption{(a) Surface morphology of the as-pyrolyzed film; (b) ATR-FT-IR spectra of four typical samples. "F/Ba-0", "all TFA" and "F/Ba-2" represent the sample were previously coated by the non-fluorine solution, conventional TFA-MOD solution and solution F/Ba-2, respectively. The temperature (400  or 700 $^\circ$C) after the solution name represents the quenching temperature of the sample.} \label{as-prolyzed}
\end{figure}

An as-pyrolyzed film, which was previously coated by the solution F/Ba-2, was obtained by quenching the film at the end of the 400 $^{\circ}$C stage in Figure \ref{HT}. Its typical surface morphology is shown in Figure \ref{as-prolyzed} (a) observed by OM. Typically, the surface of the as-pyrolyzed film is flat and smooth without undesirable features such as cracks, bubbles or bucklings. Thus using the solution F/Ba-2, the high-quality as-pyrolyzed film could be obtained within 2.5 h, which is much shorter than in the conventional TFA-MOD method.

The ATR-FT-IR spectrum of this sample (labled as "F/Ba-2, 400$^\circ$C") was shown in Figure \ref{as-prolyzed} (b). For comparison, the as-pyrolyzed films prepared by a non-fluorine solution (labeled as "F/Ba-0, 400$^\circ$C") and an conventional 100\% fluorine containing solution (labeled as "all TFA, 400$^\circ$C") were also measured. As expected, the characteristic IR-absorption bands of BaCO$_3$ centered at 1415, 1059 and 860 cm$^{-1}$ was detected in the non-fluorine sample but not present in the all TFA sample. The characteristic bands of BaCO$_3$ can also be identified in sample "F/Ba-2, 400$^\circ$C". However, they are much weaker than in the non-fluorine sample, which indicates that BaCO$_3$ was strongly suppressed by the fluorine content introduced in the precursor solution. It is also worthy to notice that when the as-pyrolyzed film was further heated to 700 $^\circ$C, BaCO$_3$ was totally eliminated, as indicated by the ATR-FT-IR spectrum marked with "F/Ba-2, 700 $^\circ$C".

\begin{figure}[ht]
\centering
\includegraphics[width=0.33\textwidth]{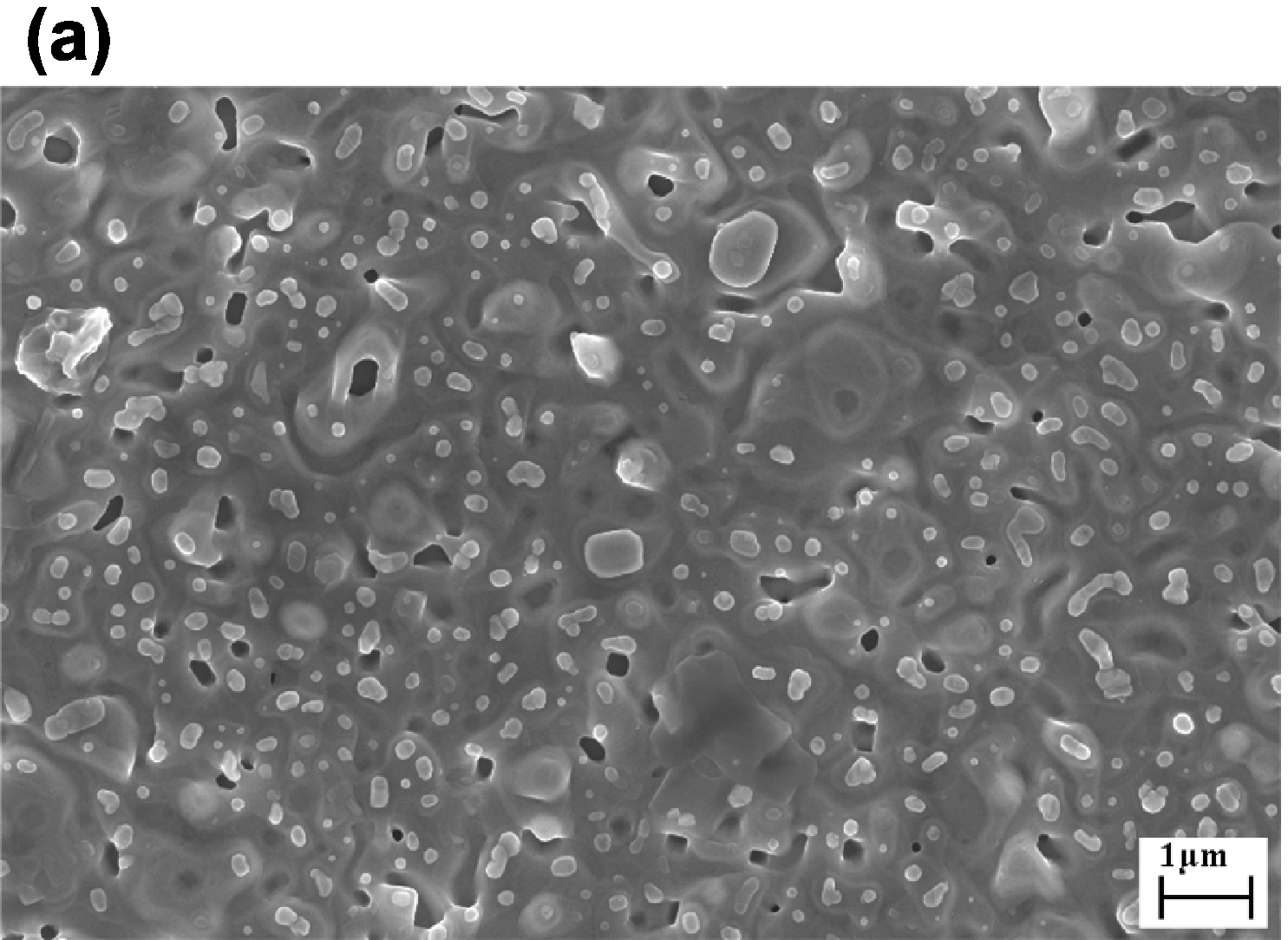}
\flushleft
\includegraphics[width=0.48\textwidth]{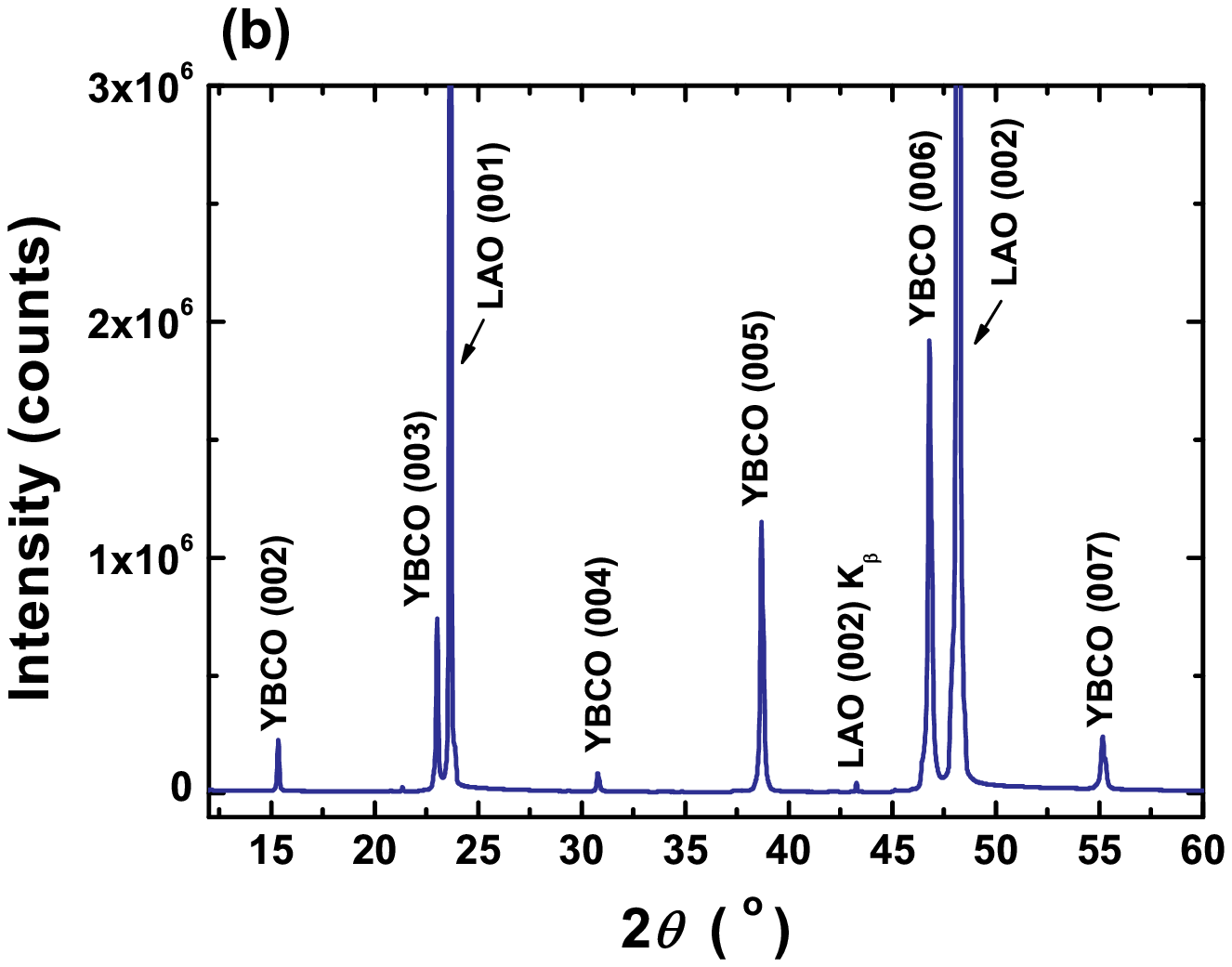}
\includegraphics[width=0.48\textwidth]{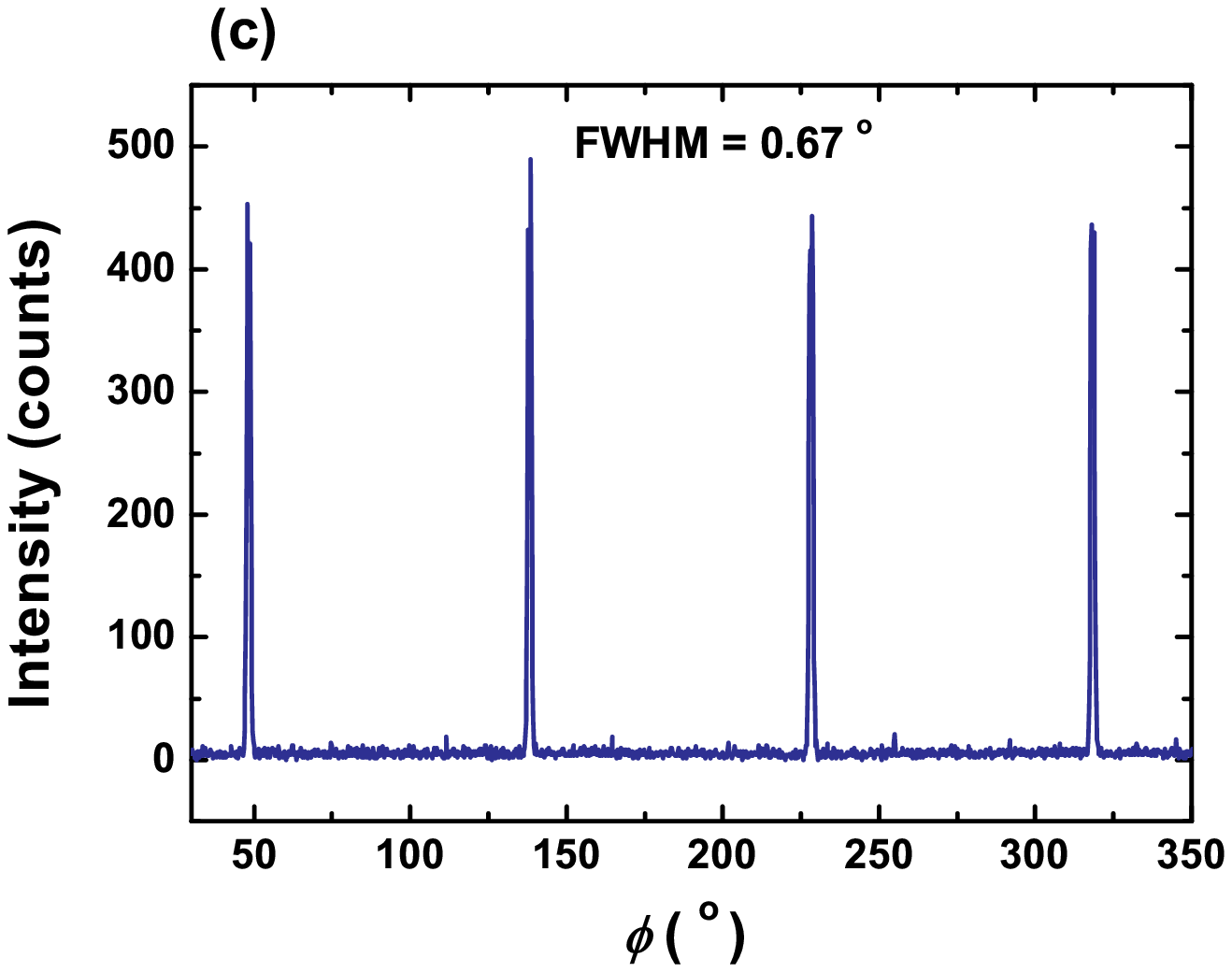}
\caption{(a) the SEM surface morphology; (b) XRD $\theta$/2$\theta$ scan pattern; (c) XRD $\phi$ scan of (102) crystallographic plane of the fully processed YBCO film.} \label{full-processed}
\end{figure}

Figure \ref{full-processed} (a) shows the SEM morphology of the fully processed YBCO film. It can be observed that this sample is composed of plate-like c-axis oriented grains with a small amount of nanoscale pores and Cu-rich surface particles. The domination of the c-axis oriented grains could be confirmed by XRD $\theta$/2$\theta$ characterization, as shown in Figure \ref{full-processed}(b). The in-plane texture of the film was examined by $\phi$ scan of the YBCO (102) crystal plane. As illustrated in Figure \ref{full-processed}(c), four sharp peaks, with the average full width at half-maximum (FWHM) value of 0.67 $^{\circ}$, corresponded to a typical four-fold symmetry of well textured YBCO.

\begin{figure}
\centering
\includegraphics[width=0.48\textwidth]{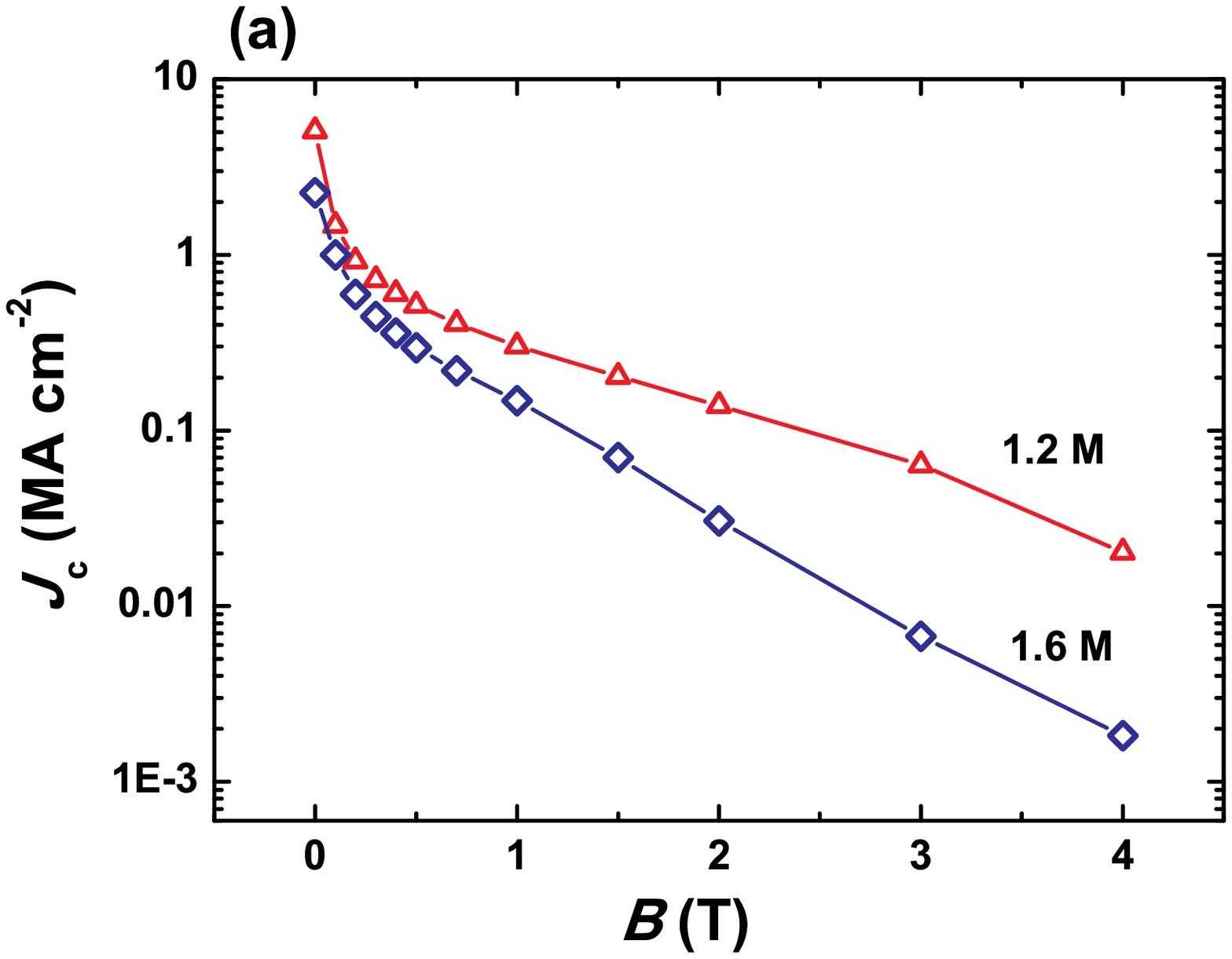}\\
\includegraphics[width=0.48\textwidth]{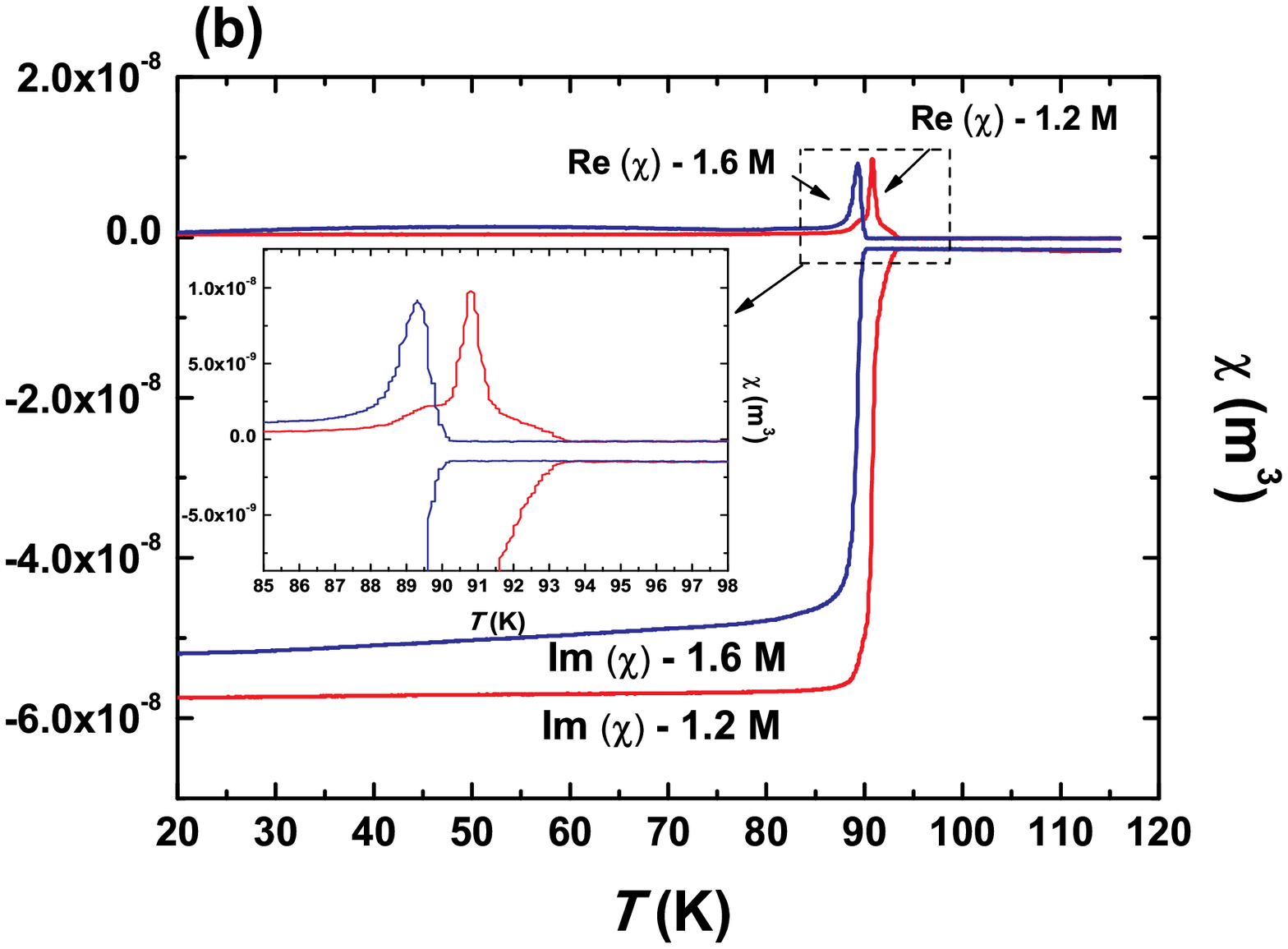}
\caption{(a) The $J_\mathrm{c}$-$B$ curves of the YBCO films coated by the solution F/Ba-2-1.2M and F/Ba-2-1.6M in the magnetic field ($B$) perpendicular to the LAO substrate at 77 K.(b) The $\chi$-$T$ curves of these two YBCO films, where Re ($\chi$) and Im ($\chi$) are the real and imaginary part of $\chi$, respectively.} \label{Jc-and-Tc}
\end{figure}

The thickness of the fully processed YBCO films prepared by solution F/B-2-1.2M and F/B-2-1.6M are 200 nm and 450 nm, respectively. As shown in Figure \ref{Jc-and-Tc} (a), at 0 T, the former one has a $J_\mathrm{c}$ of 5.1 MA cm$^{-2}$ and the later one has a $J_\mathrm{c}$ of 2.3 MA cm$^{-2}$. When the applied magnetic field $B$ was increased to 3 T, their $J_\mathrm{c}$ decreased rapidly to 0.1 and 0.01 MA cm$^{-2}$, respectively. As shown in Figure \ref{Jc-and-Tc} (b), the on-set $T_\mathrm{c}$ of the film prepared by the solution F/Ba-2-1.2M is as high as 93.5 K. Although the on-set $T_\mathrm{c}$ of the thicker film prepared by the solution F/Ba-2-1.6M is obviously lower, its value is above 90 K. Based on the results of this section, it is feasible for fabricating high-performance YBCO film with our novel low-fluorine solution.

\subsection{In-situ study of gaseous by-products during the pyrolysis step}
In the studies of the conventional 100\% fluorine containing TFA-MOD solution, fluorine containing gaseous by-products were always detected during the pyrolysis step, indicating the loss of F atoms \cite{Llordes2010CM,Mosiadz2011Ba,Mosiadz2012Cu,Mosiadz2012Y}. In this study, the fluorine content in the precursor solution was approximately the minimum requirement for the complete conversion of Ba content to BaF$_2$. Therefore, investigating whether or not fluorine content loss occurred during the pyrolysis step, which is due to the generation of fluorinated gaseous by-product(s), is necessary.

\begin{figure}[h]
\includegraphics[width=0.48\textwidth]{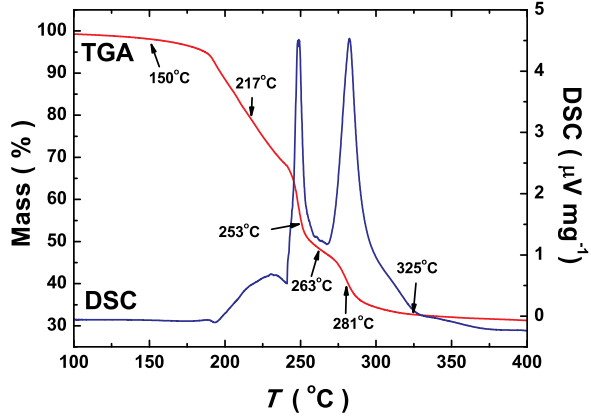}
\caption{DSC and TGA measurements of the bulk gel samples prepared by the solution F/Ba-2. A humid oxygen atmosphere and a heating rate of 3 $^\circ$C min$^{-1}$ were used.} \label{DSC-TGA}
\end{figure}

\begin{figure}[ht]
\includegraphics[width=0.48\textwidth]{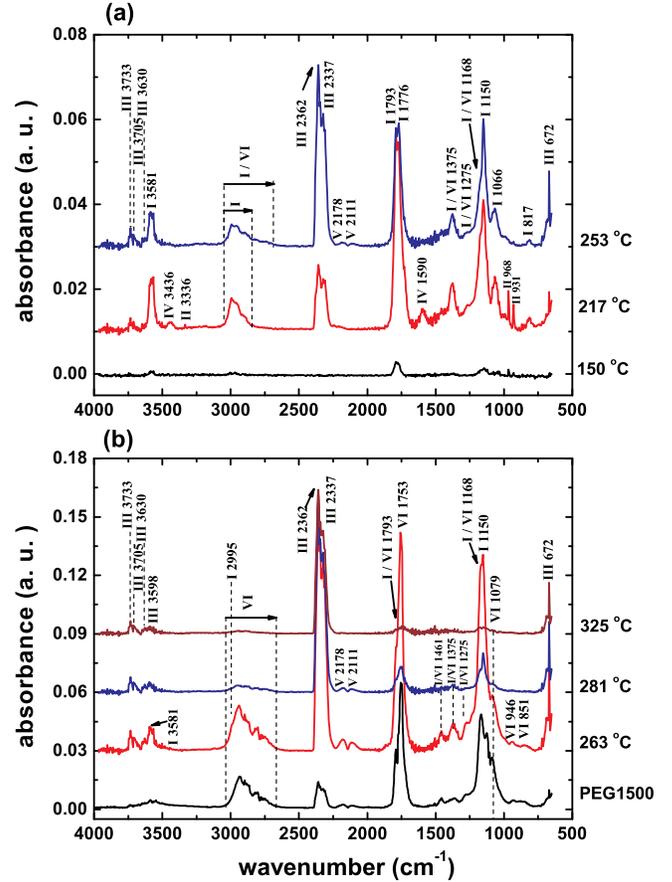}
\caption{FT-IR signals of the gaseous by-products generating at the representative temperatures shown in Figure \ref{DSC-TGA}. The indices I-VI represent CH$_3$CH$_2$COOH, NH$_3$, CO$_2$, CH$_3$CH$_2$CONH$_2$, CO and the PEG decomposition products, respectively.} \label{FT-IR}
\end{figure}

In order to identify the gaseous by-products, in-situ FT-IR spectra of them were collected simultaneously as the DSC-TGA measurement was conducted. As illustrated in the DSC curve of Figure \ref{DSC-TGA}, exothermic peaks indicating the decomposition of the PEG additive and metal salts could be observed between 190 and 325 $^{\circ}$C. The TGA curve could be divided into 6 typical regions, as labeled by their representative temperatures in Figure \ref{DSC-TGA}. The FT-IR spectra of the gaseous by-products collected at these representative temperatures are shown in Figures \ref{FT-IR}(a) and (b). Unless expressly referenced, the bands were assigned according to the FT-IR data from NIST Chemistry Web Book \cite{NIST}. Some unidentified bands (VI) were recognized to be the decomposition products of PEG1500, according to the FT-IR measurement of pure PEG1500 as shown in Figure \ref{FT-IR}(b).

At 150 $^{\circ}$C, only weak signals due to the presence of gaseous propionic acid (I) and ammonia (II) that was previously trapped in the gel matrix could be detected. No signals of CO$_2$ (III), which could indicate the decomposition of organic compounds \cite{Llordes2010CM,Matsuda2008JMR}, were detected till 217 $^{\circ}$C. At 217 $^{\circ}$C, the emerging absorption bands at 1599 cm$^{-1}$ and 3442 cm$^{-1}$ were respectively assigned to the scissoring vibration and the symmetric stretching vibration of N-H in propanamide (IV) \cite{Basiuk2000JAAP,Buczek2011JMM}. Between 253 $^{\circ}$C and 281 $^{\circ}$C, the characteristic bands corresponding to CO (V) and the decomposition products of PEG1500 (VI) appeared. On the other hand, signals of CO$_2$ (III) were much stronger, indicating that more organic compounds began to decompose. When the temperature was raised to 325 $^{\circ}$C, only weaker CO$_2$ signals could be observed, which implied that the decomposition reactions came to the end.

Considering the possible fluorine containing gaseous by-products, the characteristic bands of CHF$_3$ (1153 and 1378 cm$^{-1}$ \cite{NIST}) nearly overlapped with those of propionic acid (I); thus, determination of its presence needs other techniques like mass spectrometry. However, other fluorine containing gaseous by-products reportedly observed during the decomposition of TFA salts \cite{Llordes2010CM,Mosiadz2011Ba,Mosiadz2012Cu,Mosiadz2012Y}, including CF$_3$COOH (1200, 1123, 1830 cm$^{-1}$), CF$_3$COF (1097, 1200, 1333, 1897 cm$^{-1}$), COF$_2$(1957, 1928, 1269, 617 cm$^{-1}$) \cite{Berney1972,Craig1988}, HF (3920, 3877, 3833, 3788 cm$^{-1}$), C$_2$F4 (1338, 1186 cm$^{-1}$), C$_2$F$_6$ (1250, 1115 cm$^{-1}$) \cite{Liang1996}, were not found in our study.

\subsection{Intermediate phase evolution between the pyrolysis and crystallization step}

\begin{figure}
\includegraphics[width=0.48\textwidth]{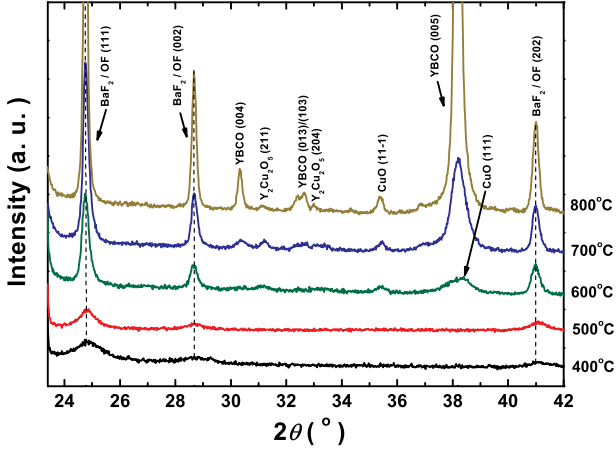}
\caption{XRD $\theta/$2$\theta$ scan patterns of films quenched at 400 - 800 $^\circ$C. } \label{phase_evolution}
\end{figure}

To study the evolution of intermediate phases between the pyrolysis and crystallization step, film samples previously quenched at 400-800 $^{\circ}$C were characterized by XRD $\theta/$2$\theta$ scan, as shown in Figure \ref{phase_evolution}. For samples quenched at 400 and 500 $^{\circ}$C, only broad peaks due to nanocrystalline BaF$_2$ could be detected. When the temperature increased to 600 $^{\circ}$C, peaks corresponding to the Y$_2$Cu$_2$O$_5$ and CuO phases emerged, and the crystallinity of the BaF$_2$ markedly improved. At the temperature higher than 700 $^{\circ}$C, the crystallinity of the BaF$_2$ phase was further enhanced, and the signal of YBCO could be detected. Peaks at the positions of BaF$_2$ may also be attributed to its partially oxidized form (denoted as OF), which is difficult to distinguish from BaF$_2$ in the XRD measurements \cite{Gazquez2006,Wesolowski2007}. BaCO$_3$(2$\theta$ $\sim$ 23.9$^{\circ}$), which could impair the superconducting performance of the resulting YBCO films \cite{PARMIGIANI1987PRB}, was not detected in any of the samples.

\begin{figure}
\includegraphics[width=0.48\textwidth]{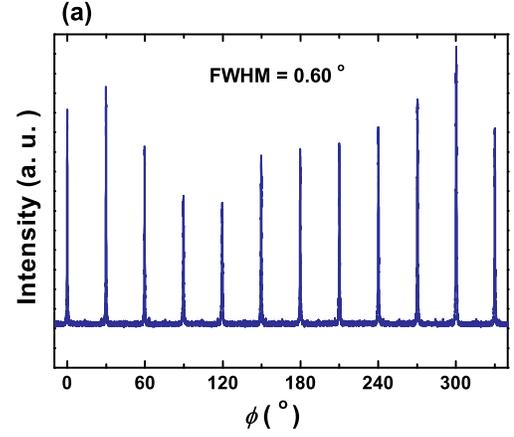}
\includegraphics[width=0.48\textwidth]{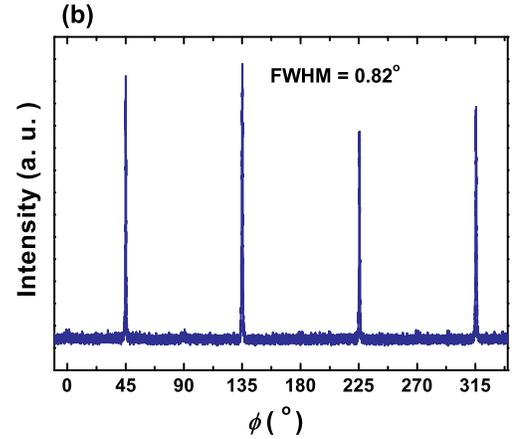}
\caption{The $\phi$ scan of the (202) crystallographic plane of (a) (111) oriented OF and (b) (001) oriented OF phases in the film quenched at 800 $^\circ$C.} \label{OF-texture}
\end{figure}

Figures \ref{OF-texture} (a) and (b) show the $\phi$ scan patterns of the (202) crystallographic plane of the (111) and (001) oriented OF phases. It can be clearly observed that the former one has a twelve-fold symmetry with an average FWHM of 0.60$^\circ$ while the first peak centered at 0$^\circ$, and the later one has a four-fold symmetry with an average FWHM of 0.82$^\circ$ while the first peak centered at 45$^\circ$. Thus it could be inferred that these two OF phases are highly textured, exhibiting the epitaxial relationships with the (001)-LAO substrate: (001)OF//(001)LAO, [110]OF//[100]LAO and (111)OF//(001)LAO, [110]OF//[100]LAO. Such relationships are in agreement with the study of the conventional TFA-MOD method \cite{Gazquez2006}.

Comparing this work with the results of previously reported MOD studies that use a conventional TFA-MOD or low-fluorine precursor solution \cite{Armenio2011,Chen2012SuST,Zalamova2006CM}, two differences in the intermediate phase evolution could be observed. First, Cu- and Y-related phases were not detected at temperatures under 600 $^{\circ}$C. We suppose it is attributed to the amorphous characteristics of these phases. Second, no evident peak shift was observed in the peaks of BaF$_2$ (or OF). This result indicates that YF$_3$, which always forms a solid solution with BaF$_2$ \cite{Zalamova2010SuST}, is not involved in the intermediate phase evolution in our study.

\section{Discussion}

Based on the ATR-FT-IR characterization of BaCO$_3$ and the intermediate phases found by the XRD $\theta/$2$\theta$ scan (Figure \ref{phase_evolution}), it could be inferred that nearly all of YBCO forms via reaction (\ref{YBCO conversion}), which has also been reported in the study of the conventional TFA-MOD method \cite{Smith1999IEEE,Zalamova2010SuST}, the ex-situ "BaF$_2$ process" \cite{Wong-Ng2004SuST} and some low-fluorine MOD studies \cite{Tokunaga2004PC,Chen2012SuST}. The small amount of BaCO$_3$, which was detected in the as-pyrolyzed film and then disappeared at 700 $^\circ$C, we propose may proceed two possible paths in the further heat treatment: converting to be a part of the OF phase and therefore involved in reaction (\ref{YBCO conversion}), or producing YBCO via the reaction proposed in the study of non-fluorine precursor solution \cite{Chu1993JMR,Xu-dissertation}.

\begin{flalign}\label{YBCO conversion}
&\begin{aligned}
&\frac{1}{2}\textrm{Y}_2\textrm{Cu}_2\textrm{O}_5 + 2\textrm{BaF}_2 + 2\textrm{CuO} + 2\textrm{H}_2\textrm{O}(g)\\
&\longrightarrow \textrm{YBa}_2\textrm{Cu}_3\textrm{O}_{6.5} + 4\textrm{HF}(g)
\end{aligned}&
\end{flalign}

It could be also inferred that BaF$_2$ formed prior to YF$_3$ and CuF$_2$, according to the above study. Some chemical equilibrium analyses were carried on to investigate this formation priority, which is equivalent to this proposition: once YF$_3$ or CuF$_2$ coexists with Ba-related phases other than BaF$_2$ (usually BaCO$_3$, Ba(OH)$_2$ or BaO), they will keep reacting to form BaF$_2$ till one of them runs out. Thus potential chemical reactions, as shown in reactions (\ref{BaCO3-YF3})-(\ref{BaO-CuF2}), were studied.

\begin{flalign}\label{BaCO3-YF3}
&\textrm{BaCO}_3 + 2/3\textrm{YF}_3 \rightleftharpoons \textrm{Ba}\textrm{F}_2 + 1/3\textrm{Y}_2\textrm{O}_3 + \textrm{CO}_2 (g)&
\end{flalign}
\begin{flalign}\label{BaCO3-CuF2}
&\textrm{BaCO}_3 + \textrm{CuF}_2 \rightleftharpoons \textrm{Ba}\textrm{F}_2 + \textrm{CuO} + \textrm{CO}_2 (g)&
\end{flalign}
\begin{flalign}\label{Ba(OH)2-YF3}
&\textrm{Ba(OH)}_2 + 2/3\textrm{YF}_3 \rightleftharpoons \textrm{Ba}\textrm{F}_2 + 1/3\textrm{Y}_2\textrm{O}_3 + \textrm{H}_2\textrm{O} (g)&
\end{flalign}
\begin{flalign}\label{Ba(OH)2-CuF2}
&\textrm{Ba(OH)}_2 + \textrm{CuF}_2 \rightleftharpoons \textrm{Ba}\textrm{F}_2 + \textrm{CuO} + \textrm{H}_2\textrm{O} (g)&
\end{flalign}
\begin{flalign}\label{BaO-YF3}
&\textrm{BaO} + 2/3\textrm{YF}_3 \rightleftharpoons \textrm{Ba}\textrm{F}_2 + 1/3\textrm{Y}_2\textrm{O}_3&
\end{flalign}
\begin{flalign}\label{BaO-CuF2}
&\textrm{BaO} + \textrm{Cu}\textrm{F}_2 \rightleftharpoons \textrm{Ba}\textrm{F}_2 + \textrm{CuO}&
\end{flalign}

The standard molar Gibbs free energy change ($\Delta G_T^{\rm \circ}$) of a chemical reaction could be calculated by equation (\ref{Gibbs change}) \cite{Klotz2008}, where the $\nu$ represents the stoichiometric coefficient, and the $\Delta G_{f,T}^{\circ}$ is the standard molar Gibbs energy for the formation of a substance involved in the reaction at the temperature of "$T$" K. The $\Delta G_{f,T}^{\circ}$ data \cite{Barin1995} of the concerned phases are summarized in Table 2. Thus $\Delta G_T^{\circ}$ values for each reaction at 700-1000 K (i.e., 427-727 $^{\circ}$C) could then be obtained and summarized in Table 3.

\begin{flalign}\label{Gibbs change}
&\Delta G_T^{\circ} = \Sigma \nu \Delta G_{f,T}^{\circ}(products) - \Sigma \nu \Delta G_{f,T}^{\circ}(reactants)&
\end{flalign}
\begin{flalign}\label{equilibrium-pressure}
&\Delta G_T^{\circ} = \textrm{R}T\textrm{ln}K_{\textrm{p}} = \textrm{R}T\textrm{ln}(P_{\textrm{e}, T}/1 \ \textrm{atm})&
\end{flalign}

\begin{table}[htbp]\label{table2}
\tabcolsep 0pt
\caption{The molar standard Gibbs free energy for the formation ($\Delta G_{i,f,T}^{\circ}$, kJ mol$^{-1}$) of concerned phases at 700-1000 K \cite{Barin1995}.} \vspace*{-12pt}
\begin{center}
\def\temptablewidth{0.45\textwidth}
{\rule{\temptablewidth}{1pt}}
\begin{tabular*}{\temptablewidth}{@{\extracolsep{\fill}}ccccc}
Phase & $\Delta G_{i,f,700}$ & $\Delta G_{i,f,800}$ & $\Delta G_{i,f,900}$ & $\Delta G_{i,f,1000}$ \\
\hline
 BaCO$_3$   & -1033 & -1007& -981 & -956 \\
 BaF$_2$    & -1092 & -1076 & -1060 & -1044 \\
 BaO        & -487 & -478 & -468 & 	-458 \\
 Ba(OH)$_2$ & -746 & -722 & -697 & 	-673 \\
 Y$_2$O$_3$ & -1699 & -1671 & -1643 & -1615\\
 YF$_3$     & -1540 & -1516 & -1492 & -1468 \\
 CuO        & -92 & -83 & 	-75 & 	-66 \\
 CuF$_2$    & -431 & -416 & -402 & 	-389 \\
 CO$_2$(g)  & -395 & -396 & -396 & -396 \\
 H$_2$O(g)  & -209 & -204 & -198 & -193 \\

       \end{tabular*}
       {\rule{\temptablewidth}{1pt}}
       \end{center}
       \end{table}

\begin{table}[htbp]\label{table3}
\tabcolsep 0pt \caption{The molar standard Gibbs free energy changes ($\Delta G_T^{\circ}$, kJ mol$^{-1}$) for the potential reactions involving the conversion of BaF$_2$ at 700-1000 K.} \vspace*{-12pt}
\begin{center}
\def\temptablewidth{0.45\textwidth}
{\rule{\temptablewidth}{1pt}}
\begin{tabular*}{\temptablewidth}{@{\extracolsep{\fill}}ccccc}
Reaction & $\Delta G_{700}^{\circ}$ & $\Delta G_{800}^{\circ}$ & $\Delta G_{900}^{\circ}$ & $\Delta G_{1000}^{\circ}$ \\
\hline
(3) & 5.55 & -10.7 & -26.8 & -42.7 \\
(4) & -116 & -131 & -146 & -161 \\
(5) & -94.4 & -104 & -113 & -122 \\
(6) & -216 & -225 & -233 & -241 \\
(7) & -145 & -145 & -145 & -145 \\
(8) & -266 & -265 & -264 & -263 \\
       \end{tabular*}
       {\rule{\temptablewidth}{1pt}}
       \end{center}
       \end{table}

\begin{table}[htbp]\label{table4}
\tabcolsep 0pt
\caption{The equilibrium pressure of CO$_2$ or H$_2$O ($P_{\textrm{e}, T}$) in reactions (3)-(6) at 700-1000 K. The unit of $P_{\textrm{e}, T}$ is atm.}
\vspace*{-12pt}
\begin{center}
\def\temptablewidth{0.45\textwidth}
{\rule{\temptablewidth}{1pt}}
\begin{tabular*}{\temptablewidth}{@{\extracolsep{\fill}}ccccc}
Reaction & $P_{\textrm{e}, 700}$ & $P_{\textrm{e}, 800}$ & $P_{\textrm{e}, 900}$ & $P_{\textrm{e}, 1000}$ \\
\hline
(3) & 0.385 & 5.03 & 36.0 & 170 \\
(4) & 4.58$\times$10$^{8}$ & 3.80$\times$10$^{8}$ & 3.16$\times$10$^{8}$ & 2.64$\times$10$^{8}$ \\
(5) & 1.11$\times$10$^{7}$ & 6.42$\times$10$^{6}$ & 3.87$\times$10$^{6}$ & 2.45$\times$10$^{6}$ \\
(6) & 1.32$\times$10$^{16}$ & 4.85$\times$10$^{14}$ & 3.40$\times$10$^{13}$ & 3.81$\times$10$^{12}$ \\
       \end{tabular*}
       {\rule{\temptablewidth}{1pt}}
       \end{center}
       \end{table}

The direction of reactions (\ref{BaCO3-YF3})-(\ref{Ba(OH)2-CuF2}) depend on the partial pressure of the gaseous product (CO$_2$ or H$_2$O). Table 4 shows the equilibrium pressure of the gaseous product ($P_{\textrm{e}, T}$) in these reactions at 700-1000 K, which was calculated by equation (\ref{equilibrium-pressure}) \cite{Alberty1987} (where $K_{\textrm{p}}$ is the equilibrium constant). Take reaction (\ref{BaCO3-YF3}) at 700 K as an example. Only when the partial pressure of CO$_2$ is above 0.385 atm, this reaction goes towards the left. However, in our experiment, the atmosphere is the flowing wet O$_2$/N$_2$ mixture and therefore the equilibrium pressure of CO$_2$ will never retain such a high value. Thus if there are BaCO$_3$ and YF$_3$, they will keep reacting to form BaF$_2$ till one of them runs out. About reactions (\ref{BaCO3-CuF2})-(\ref{Ba(OH)2-CuF2}), we can draw same conclusions no matter the gaseous product is CO$_2$ or H$_2$O. Reactions (\ref{BaO-YF3}) and (\ref{BaO-CuF2}), which do not involve gaseous reactant or product, have $\Delta G_{T}^{\circ}$ much lower than 0, indicating BaO will also convert to BaF$_2$ in a thermodynamically stable state. According to the analyses above, the formation priority of BaF$_2$ over YF$_3$ and CuF$_2$ is verified.

\section{Summary}

In this study, a novel low-fluorine precursor solution for MOD-YBCO fabrication was developed. The TFA amount in this solution was only about 10.3\% of that used in the conventional TFA-MOD solution, corresponding to a starting F/Ba mole ratio of 2. ATR-FT-IR revealed that BaCO$_3$ was remarkably suppressed in the as-pyrolyzed film and eliminated at 700$^\circ$C. Thus YBCO films with $J_\mathrm{c}$ of 5.1 MA cm$^{-2}$ (77 K, 0 T, 200 nm thickness) could be obtained on LAO substrate. In-situ FT-IR spectra showed that no obvious fluorinated gaseous by-products were detected in the pyrolysis step, which indicated that all of the F atoms might remain in the film as fluorides. XRD $\theta$/2$\theta$ patterns showed that BaF$_2$ (or OF), but neither YF$_3$ nor CuF$_2$, was detected in the films quenched at 400 - 800 $^\circ$C, which together with Y$_2$Cu$_2$O$_5$, CuO and H$_2$O could react to form YBCO . The formation priority of BaF$_2$ over YF$_3$ and CuF$_2$ was interpreted by the chemical equilibrium of the potential reactions. The 10.3\% fluorine content used in the precursor solution of this study was lower than that in all the reported low-fluorine studies. Therefore, this study could enlarge the synthesis window of the precursor solution for MOD-YBCO fabrication, and help to study the influence of the fluorine content in the precursor solution continuously and systematically, which may be beneficial for the YBCO thick film fabrication. Such application of this method will be conducted in our future research.


\section*{Acknowledgements}
This work was supported by the National Natural Science Foundation of China (51202124), China Postdoctoral Science Foundation (2013M530615), the Fundamental Research Program of Shenzhen (JCYJ20120614193005764), and the Danish Agency for Science, Techonology and Innovation (09-062997). The authors thank Prof. Niels Hessel Andersen from physics department of technical university of Denmark for the great help in measuring the superconducting properties.



\newpage
\begin{thebibliography}{10}
\expandafter\ifx\csname url\endcsname\relax
  \def\url#1{{\tt #1}}\fi
\expandafter\ifx\csname urlprefix\endcsname\relax\def\urlprefix{URL }\fi
\providecommand{\eprint}[2][]{\url{#2}}

\bibitem{Gupta1988}
Gupta A, Jagannathan R, Cooper E, Giess E, Landman J and Hussey B {1988} {\em
  {Appl. Phys. Lett.}\/} {\bf {52}} {2077--2079}

\bibitem{MCINTYRE1990JAP}
McIntyre P~C, Cima M~J and Ng M~F {1990} {\em {J. Appl. Phys.}\/} {\bf {68}}
  {4183--4187}

\bibitem{Smith1999IEEE}
Smith J~A, Cima M~J and Sonnenberg N {1999} {\em {IEEE Trans. Appl.
  Supercond.}\/} {\bf {9}} {1531--1534}

\bibitem{Obradors2006SuST}
Obradors X, Puig T, Pomar A, Sandiumenge F, Mestres N, Coll M, Cavallaro A,
  Roma N, Gazquez J, Gonzalez J, Castano O, Gutierrez J, Palau A, Zalamova K,
  Morlens S, Hassini A, Gibert M, Ricart S, Moreto J, Pinol S, Isfort D and
  Bock J {2006} {\em {Supercond. Sci. Technol.}\/} {\bf {19}} {S13--S26}

\bibitem{Araki2003SuST}
Araki T and Hirabayashi I {2003} {\em {Supercond. Sci. Technol.}\/} {\bf {16}}
  {R71--R94}

\bibitem{Araki2001}
Araki T, Takahashi Y, Yamagiwa K, Iijima Y, Takeda K, Yamada Y, Shibata J,
  Hirayama T and Hirabayashi I {2001} {\em {Physica C}\/} {\bf {357}}
  {991--994}

\bibitem{Tokunaga2004PC}
Tokunaga Y, Fuji H, Teranishi R, Matsuda J, Asada S, Kaneko A, Honjo T, Izumi
  T, Shiohara Y, Yamada Y, Murata K, Iijima Y, Saitoh T, Goto T, Yoshinaka A
  and Yajima A {2004} {\em {Physica C}\/} {\bf {412}} {910--915}

\bibitem{Xu2005}
Xu Y, Goyal A, Leonard K and Martin P {2005} {\em {Physica C}\/} {\bf {421}}
  {67--72}

\bibitem{Nakaoka2007}
Nakaoka K, Matsuda J, Kitoh Y, Goto T, Yamada Y, Izumi T and Shiohara Y {2007}
  {\em {Physica C}\/} {\bf {463}} {519--522} {19th International Symposium on
  Superconductivity, Nagoya, JAPAN, OCT 30-NOV 01, 2006}

\bibitem{Chen2012SuST}
Chen Y, Wu C, Zhao G and You C {2012} {\em {Supercond. Sci. Technol.}\/} {\bf
  {25}} 062001

\bibitem{Ichikawa2009PC}
Ichikawa H, Nakaoka K, Miura M, Sutoh Y, Nakanishi T, Nakai A, Yoshizumi M,
  Izumi T and Shiohara Y {2009} {\em {Physica C}\/} {\bf {469}} {1329--1331}

\bibitem{Li2007IEEE}
Li X, Rupich M~W, Kodenkandath T, Huang Y, Zhang W, Siegal E, Verebelyi D~T,
  Schoop U, Nguyen N, Thieme C, Chen Z, Feldman D~M, Larbalestier D~C,
  Holesinger T~G, Civale L, Jia Q~X, Maroni V,  and Rane M~V 2007 {\em IEEE
  Trans. Appl. Supercond.\/} {\bf 17} 3553

\bibitem{Feenstra2009IEEE}
Feenstra R, List F~A, Li X, Rupich M~W, Miller D~J, Maroni V~A, Zhang Y,
  Thompson J~R and Christen D~K 2009 {\em IEEE Trans. Appl. Supercond.\/} {\bf
  19} 3131

\bibitem{Yoshizumi2004PC}
Yoshizumi M, Seleznev I and Cima M 2004 {\em Physica C\/} {\bf 403} 191

\bibitem{Wesolowski2006PC}
Wesolowski D, Yoshizumi M and Cima M 2006 {\em Physica C\/} {\bf 450} 76

\bibitem{Gazquez2006}
Gazquez J, Sandiumenge F, Coll M, Pomar A, Mestres N, Puig T, Obradors X, Kihn
  Y, Casanove M~J and Ballesteros C {2006} {\em {Chem. Mat.}\/} {\bf {18}}
  {6211--6219}

\bibitem{Zalamova2010SuST}
Zalamova K, Pomar A, Palau A, Puig T and Obradors X {2010} {\em {Supercond.
  Sci. Technol.}\/} {\bf {23}} {014012}

\bibitem{Matsuda2008JMR}
Matsuda J, Nakaoka K, Izumi T, Yamada Y and Shiohara Y {2008} {\em {J. Mater.
  Res.}\/} {\bf {23}} {3353--3362}

\bibitem{Armenio2011}
Armenio A~A, Augieri A, Ciontea L, Contini G, Davoli I, Di~Giovannantonio M,
  Galluzzi V, Mancini A, Rufoloni A, Petrisor T, Vannozzi A and Celentano G
  {2011} {\em {Supercond. Sci. Technol.}\/} {\bf {24}} 115008

\bibitem{Wong-Ng2004SuST}
Wong-Ng W, Levin I, Feenstra R, Cook L and Vaudin M {2004} {\em {Supercond.
  Sci. Technol.}\/} {\bf {17}} {S548--S556}

\bibitem{Wu2013}
Wu W, Feng F, Shi K, Zhai W, Qu T~M, Huang R~X, Tang X, Wang X~H, Hu Q~Y,
  Grivel J~C and Han Z~H 2013 {\em Supercond. Sci. Technol.\/} {\bf 26} 055013

\bibitem{Vermeir2009SuST}
Vermeir P, Cardinael I, Baecker M, Schaubroeck J, Schacht E, Hoste S and
  Van~Driessche I {2009} {\em {Supercond. Sci. Technol.}\/} {\bf {22}} 075009

\bibitem{GYORGY1989APL}
Gyorgy E~M, Vandover R~B, Jackson K~A, Schneemeyer L~F and Waszczak J~V {1989}
  {\em {Appl. Phys. Lett.}\/} {\bf {55}} {283--285}

\bibitem{Llordes2010CM}
Llordes A, Zalamova K, Ricart S, Palau A, Pomar A, Puig T, Hardy A, Van~Bael
  M~K and Obradors X {2010} {\em {Chem. Mat.}\/} {\bf {22}} {1686--1694}

\bibitem{Mosiadz2011Ba}
Mosiadz M, Juda N~L, Hopkins S~C, Soloducho J and Glowacki B~A {2011} {\em
  {Thermochim. Acta}\/} {\bf {513}} {33--37}

\bibitem{Mosiadz2012Cu}
Mosiadz M, Juda K~L, Hopkins S~C, Soloducho J and Glowacki B~A {2012} {\em {J.
  Fluor. Chem.}\/} {\bf {135}} {59--67}

\bibitem{Mosiadz2012Y}
Mosiadz M, Juda K~L, Hopkins S~C, Soloducho J and Glowacki B~A {2012} {\em {J.
  Therm. Anal. Calorim.}\/} {\bf {107}} {681--691}

\bibitem{NIST}
{NIST Chemistry Webbook, http://webbook.nist.gov/chemistry/}

\bibitem{Basiuk2000JAAP}
Basiuk V~A and Douda J {2000} {\em {J. Anal. Appl. Pyrolysis}\/} {\bf {55}}
  {235--246}

\bibitem{Buczek2011JMM}
Buczek A, Kupka T and Broda M~A {2011} {\em {Journal of Molecular Modeling}\/}
  {\bf {17}} {2265--2274}

\bibitem{Berney1972}
Berney C~V and Cormier A~D 1972 {\em Spectrochim Acta A.\/} {\bf 28} 1813--1822

\bibitem{Craig1988}
Craig N~C 1988 {\em Spectrochim Acta A.\/} {\bf 44} 1225--1226

\bibitem{Liang1996}
Liang J, Safriet A, Briley S and Roselius M {1996} {\em {J. Fluor. Chem.}\/}
  {\bf {79}} {53--57}

\bibitem{Wesolowski2007}
Wesolowski D~E, Yoshizumi M and Cima M~J {2007} {\em {IEEE Trans. Appl.
  Supercond.}\/} {\bf {17}} {3351--3354}

\bibitem{PARMIGIANI1987PRB}
Parmigiani F, Chiarello G, Ripamonti N, Goretzki H and Roll U {1987} {\em
  {Phys. Rev. B}\/} {\bf {36}} {7148--7150}

\bibitem{Zalamova2006CM}
Zalamova K, Roma N, Pomar A, Morlens S, Puig T, Gazquez J, Carrillo A~E,
  Sandiumenge F, Ricart S, Mestres N and Obradors X {2006} {\em {Chem. Mat.}\/}
  {\bf {18}} {5897--5906}

\bibitem{Chu1993JMR}
Chu P and Buchanan R 1993 {\em J. Mater. Res.\/} {\bf 8} 2134--2142

\bibitem{Xu-dissertation}
{Xu Y} {2003} {\em {High Jc epitaxial YBa$_2$Cu$_3$O$_{7-x}$ films through a
  non-fluorine approach for coated conductor applications}\/} Ph.D. thesis
  {Dept. Chem. Mater. Eng., Univ. Cincinnati,} {Cincinnati, OH}

\bibitem{Klotz2008}
{Klotz I M} {2008} {\em {Chemical thermodynamics: basic concepts and
  methods}\/} ({New Jersey}: {John Wiley \& Sons, Inc.})

\bibitem{Barin1995}
{Barin I} {1995} {\em {Thermaochemical data of pure substances, 3rd Edition}\/}
  ({Weinheim, Germany}: {WILEY-VCH Verlag GmnH})

\bibitem{Alberty1987}
{Alberty R A} {1987} {\em {Physical Chemistry, 7th Edition}\/} ({New York,
  USA}: {John Wiley \& Sons, Inc.})

\end{thebibliography}
\end{document}